# BIAS IN VELOCITY FIELD RECOVERIES

A. NEWSAM[1], J.F.L. SIMMONS[1], M.A. HENDRY[2]
[1] *University of Glasgow, Glasgow, UK*
[2] *University of Sussex, Falmer, Brighton, UK*

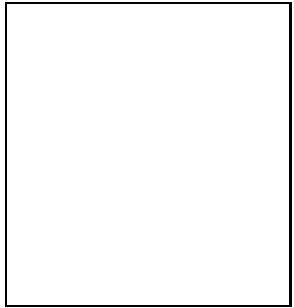

**Abstract**

We investigate the effect of using different distance estimators on the recovery of the peculiar velocity field of galaxies using POTENT. An inappropriate choice of distance estimator will give rise to spurious flows. We discuss methods of minimising these biases and the levels of accuracy required of distance estimators to retrieve velocity fields to a given standard.

## 1 Introduction

Recently cosmologists have paid a great deal of attention to the problem of constructing the peculiar velocity field of galaxies. The main importance of such velocity field reconstructions lies in their implications for the large scale distribution of matter: any systematic deviations of the velocity field from quiet Hubble flow indicate density inhomogeneities, the measurement of which can place constraints upon $\Omega$.

The essential idea of POTENT ([1], [2], [4] hereafter DBF, [5]) is to derive this underlying peculiar velocity field directly from measurements of galaxy redshifts. The fundamental assumption is that the peculiar velocity field can be represented by a gradient of a scalar potential function. It then follows that this potential can be derived by taking the line integral of the redshifts along a radial path, provided of course that the galaxy distances are accurately known and that the galaxies about which we have this information are sufficiently densely distributed.

However the source of our difficulties lies in our failure to meet these modest requirements. Although the number of galaxies appearing in redshift surveys has increased enormously [10], there are still large regions of the sky where data are extremely sparse. Evidently in these regions there is little hope of reconstructing reliable velocity fields, without making further physical assumptions based on dynamics or N body simulations.

The peculiar velocity of a galaxy is found by subtracting $Hr$ from its observed redshift and so the accuracy of this inferred peculiar velocity will depend on the accuracy of the estimated distance of the galaxy. The contentious issue of distance estimation has been discussed in the context of velocity field reconstruction by a number of authors ([8], [7] hereafter LS, DBF, [11], [6]) and centres around the so called Malmquist effect [9]. Despite the claim by LS to have solved this long standing difficulty in accounting for the effects of selection on distance estimation, it seems to us that there is still considerable confusion even over the nature of the Malmquist effect, let alone its correction. However, we discuss this problem elsewhere in these proceedings [6] and shall here concentrate on the effect on POTENT.

In this paper we repeat the POTENT calculations of DBF, applying the homogeneous and inhomogeneous corrections to demonstrate the radical differences in reconstructed velocity fields. We carry this out on both simulated data and on data obtained from [10] and [3] and investigate the claim that the inhomogeneous Malmquist correction to distance estimates is able to eradicate bias in the reconstructed smoothed velocity fields. We also attempt to quantify the errors in the components of smoothed velocity field obtained using POTENT and determine how these depend on the accuracy of the distance estimators.

## 2 How Potent Works

Full descriptions of the POTENT method are given by its originators in DBF, but it is convenient to summarise here the basic ideas of the method so that we can subject some of the assumptions to closer scrutiny.

The key idea of POTENT is to write

$$v(r, \theta, \phi) = -\nabla \Phi(r, \theta, \phi) \tag{1}$$

and hence obtain $\Phi(r, \theta, \phi)$ from a suitable line integral. Taking a radial path will involve only the radial components of the velocity fields, and hence

$$\Phi(r, \theta, \phi) = -\int_0^r v_r dr \tag{2}$$

where $v_r$ is given by the redshifts.

The problem in carrying out this radial line integral is that

**(i)** the radial component of the peculiar velocity can only be obtained at those points where galaxies are found and

**(ii)** these peculiar velocities are only estimates, and rely on the estimation of the distances to galaxies thus

$$\hat{\mathbf{u}}_\mathbf{i} = cz_i - H_0 \hat{\mathbf{r}}_\mathbf{i} \tag{3}$$

Here and in future a ˆ will indicate an estimator. We shall also indicate a statistical variable by bold face, and its value by the corresponding regular letter.

To cope with sparseness DBF adopt the method of tensor window functions. This method obtains a smoothed peculiar velocity field, $\vec{v}$, at every spatial point, by best fitting to the estimated radial components of the peculiar velocity $\hat{\mathbf{u}}_\mathbf{i}$. This is done by minimising

$$\sum_{i=1}^{N} (\hat{\vec{v}} . \vec{e}_r(\hat{\vec{r}}_i) - u_i)^2 W(\vec{r}, \hat{\vec{r}}_i) \tag{4}$$

where $W(\vec{r}, \vec{r}_i)$ is a chosen window function, $\vec{r}_i$ is the position vector of the $i^{th}$ galaxy, and there are $N$ galaxies is the catalogue. $\vec{e}_r(\vec{r}_i)$ is the unit vector in the radial direction of the $i^{th}$ galaxy at position $\vec{r}_i$. (N.B. $\vec{r}_i$ and $\hat{\vec{r}}_i$ will be in the same direction.)

There are several crucial points to observe in this procedure.

(i) Even in the absence of distance errors the smoothed peculiar velocity field will not be exact. DBF consider the case where an input smooth peculiar velocity field is sampled at various points corresponding to galaxies, and reconstructed using procedure given by equation (4). This rederived field will be subject to 'sampling gradient bias', which will be particularly acute where the galaxies are sparse. A good choice of window function will help to minimise this effect, but for spatially inhomogeneous samples the effect cannot be removed everywhere.

(ii) If noise is introduced into the smoothed input field, and/or distance estimates are subject to errors, the mean retrieved field obtained through smoothing will in general not be equal to the input smooth field. In other words the retrieved smoothed field will be biased. DBF call this Malmquist bias, in distinction to the sampling gradient bias, and have attempted to remove it by applying the homogeneous [8] Malmquist corrections to the distance estimates. They do not appear to have used the inhomogeneous correction proposed by LS. We shall discuss the justification of this procedure below.

## 3 Bias in Distance Estimators

There have been in the last decade rapid developments in the techniques of determining distances of galaxies. Most of these depend on finding an observable, which we denote generically by P, that correlates strongly with the absolute magnitude of the galaxy, or with the absolute diameter. Thus by measuring P, one can infer an estimate of the absolute magnitude or diameter. The distance to the galaxy may then be determined from the observed apparent magnitude or apparent diameter. The Tully-Fisher and $D_n - \sigma$ are examples of distance indicators based on this principle. Two difficulties arise, in addition to the more obvious problems of correcting for absorption, spacetime curvature effects etc. Firstly it is necessary to find the correlation coefficients and zero point from a cluster of galaxies at a known distance. Secondly, one must account for the effects of selection both when one carries out the calibration from the cluster of galaxies, and when estimating the distance of any given galaxy.

Consider the simplest case where only the apparent magnitude, $m$, is observed. For any given galaxy of absolute magnitude, $M$ and distance $r$ (in Mpc), we have

$$\log r = 0.2(m - M - 25) \tag{5}$$

Consider now a galaxy that is selected from a fictitious population of galaxies all at distance $r_0$ with random absolute magnitude. The mean absolute magnitude is taken to be $M_0$. Only when we stipulate the absolute magnitude of the galaxy is the distance estimator well defined. Thus an obvious choice of estimator for $\log r$ is

$$\hat{\omega} = \hat{\log} \mathbf{r} = 0.2(\mathbf{m} - M_0 - 25) \tag{6}$$

Since $\mathbf{M}$ is sampled from the luminosity function, both $\mathbf{M}$ and $\mathbf{m}$ should be considered statistical variables. We shall denote the expectation or mean value by $E$, where the averaging is over the observable population, having taken selection into account. It is easy to show that $E(\mathbf{M}|r_0, M_0) \neq M_0$, and hence that

$$E(\hat{r}|r_0, M_0) \neq r_0 \text{ and } E(\hat{\omega}|\omega_0, M_0) \neq \omega_0 \tag{7}$$

Thus these estimators of both $r_0$ and $\omega_0$ are, in the statistical sense of the word, biased.

In the case where there are two observables, it is possible to construct an estimator that is linear in both which is unbiased. Thus we may write

$$\hat{\omega} = 0.2(\mathbf{m} - \hat{\mathbf{M}} - 25) \tag{8}$$

where the estimated value of $M$ is obtained from the observed value of $P$. Of course to be effective $\mathbf{M}$ and $\mathbf{P}$ must be highly correlated. What is interesting is that if galaxies are subject to selection on apparent magnitude an unbiased estimator of $\log r$ is obtained by taking $\hat{\mathbf{M}}$ from the regression line of $\mathbf{P}$ on $\mathbf{M}$ rather than the regression line of $\mathbf{M}$ on $\mathbf{P}$ [6]. The former we shall call a P-on-M estimator (giving $\hat{\omega}_{\mathbf{PM}}$), and the latter a M-on-P estimator ($\hat{\omega}_{\mathbf{MP}}$). Thus one should use the P-on-M estimator if one wants an unbiased estimator of $\log r_0$.

On the assumption that the intrinsic joint distribution of $\mathbf{P}$ and $\mathbf{M}$ is bivariate normal, $\hat{\omega}_{\mathbf{PM}}$ will be gaussian and unbiased under magnitude selection. It can be easily shown that $\hat{\omega}_{\mathbf{MP}}$ is also unbiased in the absence of selection on apparent magnitude. Moreover,

$$\hat{\mathbf{r}} = 10^{\frac{1}{2}(\log_e 10)^2} 10^{\hat{\omega}_{\mathbf{PM}}} \tag{9}$$

is an unbiased estimator of $r_0$, where $\sigma^2$ is the variance of $\hat{\omega}_{\mathbf{PM}}$.

A lot of confusion has been created by the different meanings that have been attached to the term bias (even leaving aside the cosmological biasing parameter $b$). Our viewpoint, essentially frequentist, is that the actual distance, $r_0$, of an observed galaxy is an unknown parameter (or unknown state of nature in statistical parlance) and not a statistical variable. An unbiased estimator will on average yield the value $r_0$ whatever the value of $r_0$ really is.

[8], and LS take a different view, assigning a prior probability distribution to $r_0$, based on an assumed spatial density distribution and selection function. Their argument is that the latter two factors determine the probability that the galaxy appears in the catalogue or survey. Following the measurement of $\hat{r}$, which is based on such observables as the apparent magnitude and line width, the posterior distribution of $r_0$ will differ from its prior. Thus $r_0$ is treated as a statistical variable. Their definition of an estimator being unbiased is that for such an estimator

$$E(\mathbf{r_0}|\hat{\mathbf{r}} = s) = \int \mathbf{r_0} p(\mathbf{r_0}|\hat{\mathbf{r}} = s) d\mathbf{r_0} = \frac{\int \mathbf{r_0} p(\hat{\mathbf{r}} = s|\mathbf{r_0}) P(\mathbf{r_0}) d\mathbf{r_0}}{\int p(\hat{\mathbf{r}} = s|\mathbf{r_0}) P(\mathbf{r_0}) d\mathbf{r_0}} = s \tag{10}$$

For this reason [8] apply their so-called homogeneous Malmquist correction to $\hat{r}$. Whereas [8] assume a number density of galaxies that varies as $r^\alpha$, LS construct an inhomogeneous correction using the sample distribution to estimate the prior probability distribution. Thus in principle their correction applies for arbitrary prior distributions, but it does rest on other rather dubious assumptions.

Which definition of bias one should take is not an entirely clear cut question. We shall discuss this elsewhere [6], and restrict ourselves to the remark that whichever approach one uses, it is important to be consistent.

[8] assume that $\hat{\omega}$ ($l_e$ in their notation) has a gaussian distribution with mean $\omega_0$ ($l$). (This so-called 'raw' log distance is then corrected to $l_e + \alpha\Delta^2$ and 'raw' distance to $r_e(1 + \alpha\Delta^2)$, $\Delta^2$ being the variance of $\ln D_n$ for given $\sigma$ (velocity dispersion) and $\alpha$ the exponent of the power law distribution.) However, by regressing $\mathbf{M}$ on $\mathbf{P}$, rather than $\mathbf{P}$ on $\mathbf{M}$, Lynden-Bell et al appear to have chosen $l_e = \hat{\omega}_{\mathbf{MP}}$ which, under magnitude/apparent diameter selection will be neither gaussian nor unbiased.

Since one has little information about how galaxies are selected to appear in a catalogue or survey, and we have no prior information about the number density of galaxies, we consider the distances, $r_{i0}$, of the galaxies to be unknown parameters. $\omega_{i0} = \log r_{i0}$ of the $i^{th}$ galaxy is estimated by taking the appropriate estimator, $\hat{\omega}_\mathbf{i}$, which we shall take to be the P-on-M estimator. Different estimators $\hat{\omega}_\mathbf{i}$ will have different distributions, depending on the actual value of $\omega_{i0}$ for the particular $(i^{th})$ galaxy. If we form the histogram of all values of $\hat{\omega}_\mathbf{i}$ for any given catalogue of galaxies then we should expect it to have a larger dispersion than the actual distances, but the expected mean will be equal to the mean of the histogram of $\omega_{i0}$.

DBF consider the $\omega_{i0}$ to be sampled from an underlying distribution of galaxies, in which case $\omega_\mathbf{i0}$ are all statistical variables. This accounts for their two levels of averaging, one over distance errors, and the other over different realisations of the catalogue.

## 4  Applications to Potent

Whether or not a distance estimator is biased is not the crucial question when attempting to correct for bias in POTENT. What is important is to construct an unbiased smoothed peculiar velocity field. POTENT attempts to construct an unbiased peculiar velocity field in the following sense:

One assumes an underlying smoothed peculiar velocity field (taken to be potential) and effective density distribution that is determined by some selection function and underlying density distribution of galaxies.

**(i)** This field is sampled at $n$ points and the galaxies taken to be at these at the corresponding actual distances $(r_{10}, r_{20}, r_{30}..r_{n0})$

**(ii)** Errors are added to these distances. A smoothed initial radial peculiar velocity field is derived using the tensor window function.

**(iii)** Hence, one obtains a potential velocity field by radial integration.

If this smoothed potential is the same as the input potential when it is averaged over all realisations of $(r_{10}, r_{20}, r_{30}..r_{n0})$ and of the distance errors, it is unbiased.

DBF attempt to prove that if one applies the homogeneous Malmquist correction to distance discussed above one does obtain peculiar velocity fields that are almost unbiased. Essentially their argument depends on making several Taylor expansions in $\epsilon_i$, and discarding terms of order 3 and above. If the errors, $\epsilon_i$, are large, as they will be at large distances, this procedure might break down. DBF resort to Monte Carlo simulations to back up their analytic treatment.

We discuss the question of homogeneous and inhomogeneous Malmquist corrections elsewhere [12]. However, we would like to briefly discuss why they appear to work for the POTENT analysis, although there seems to be no convincing proof. In this respect, the important factor is the window function. In 'interpolating' a peculiar velocity from galaxies appearing in the catalogue to a given spatial point with radial coordinate $s$, the *essential* effect of the window function is to pick out the galaxy whose *estimated* position is nearest to the prescribed point. This galaxy's actual distance could be radically different, and will depend on the spatial distribution of galaxies. By requiring that on average the actual radial coordinate of the galaxy deemed to be closest equals $s$ one would ensure also that on average the correct peculiar velocity would be ascribed to $s$. Expressed mathematically we require

$$E(\mathbf{r_0}|\hat{\mathbf{r}}_\mathbf{0} = s) = s \qquad (11)$$

Of course this will only work if galaxies are not too sparse, and if the gradient of velocity field is not too large, or the effective radius of the window function is not too wide.

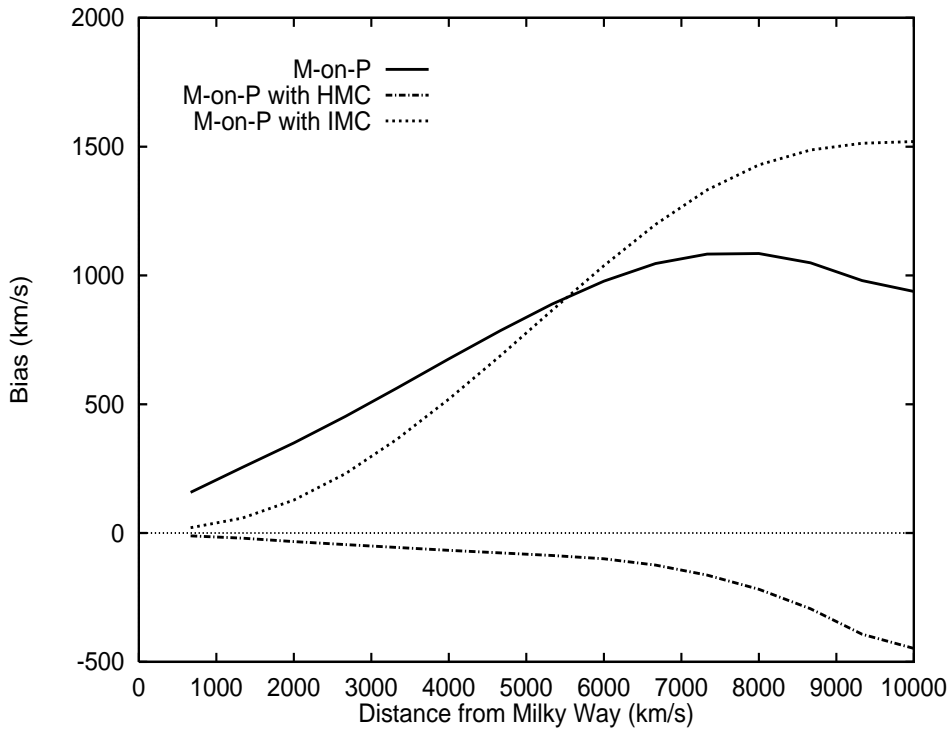

Figure 1: Bias as a function of distance for POTENT recoveries. The M-on-P regression has been used on its own, with an homogeneous Malmquist correction and with an inhomogeneous correction. The distance estimator has an error in log distance of about 10%.

## 5 Numerical Techniques

DBF make the assumption that the selection function depends only on the estimator of distance, whilst the inhomogeneous correction of LS assumes that selection depends only on the actual distance of the galaxy. For simple selection on apparent magnitude with either M-on-P or P-on-M estimators, selection will depend on P as well as $\omega_{\mathbf{MP}}$ or $\omega_{\mathbf{PM}}$. Thus neither the assumption made by DBF nor the assumption made by LS would be correctly applied to such estimators.

In this section we describe the results obtained for quiet Hubble flow when M-on-P and P-on-M are used to estimate distances from a sample of galaxies whose $\mathbf{M}, \mathbf{P}$ have a bivariate normal distribution. Our aim is to compare the reconstructs of the peculiar velocity field from POTENT using both uncorrected and corrected estimators.

Apart from the distance estimators, our analysis follows that of POTENT$_{90}$ [5], and so the results should be comparable.

To generate estimated distances of galaxies, $\mathbf{M}$ and $\mathbf{P}$ are sampled from a bivariate normal distribution and subjected to magnitude selection. We take typical values of the distribution parameters obtained for the $D_n$-$\sigma$ and Tully-Fisher relation.

## 6 Results

The effect on the reconstructed velocity field of the various combinations of distance estimators and corrections is very significant and is summarised in figures 1 and 2. These show that the biases produced for these log distance errors of about 10% are very large except for the M-on-P with homogeneous Malmquist correction (HMC) and P-on-M with inhomogeneous correction (IMC). However, we know that the HMC is invalid for two reasons. Firstly, M-on-P estimators are neither unbiased nor lognormal which are both basic assumptions of the method. Even more significant, we are considering the case of an homogeneous universe. The HMC is concerned with an homogeneous *sample* and so, as we have selection, the correction cannot expect to be valid. This is borne out by the results given in figure 3 where the log distance error has been increased to 15%. Now the corrected M-on-P is far from adequate, whereas the P-on-M with IMC still seems to be reasonably effective.

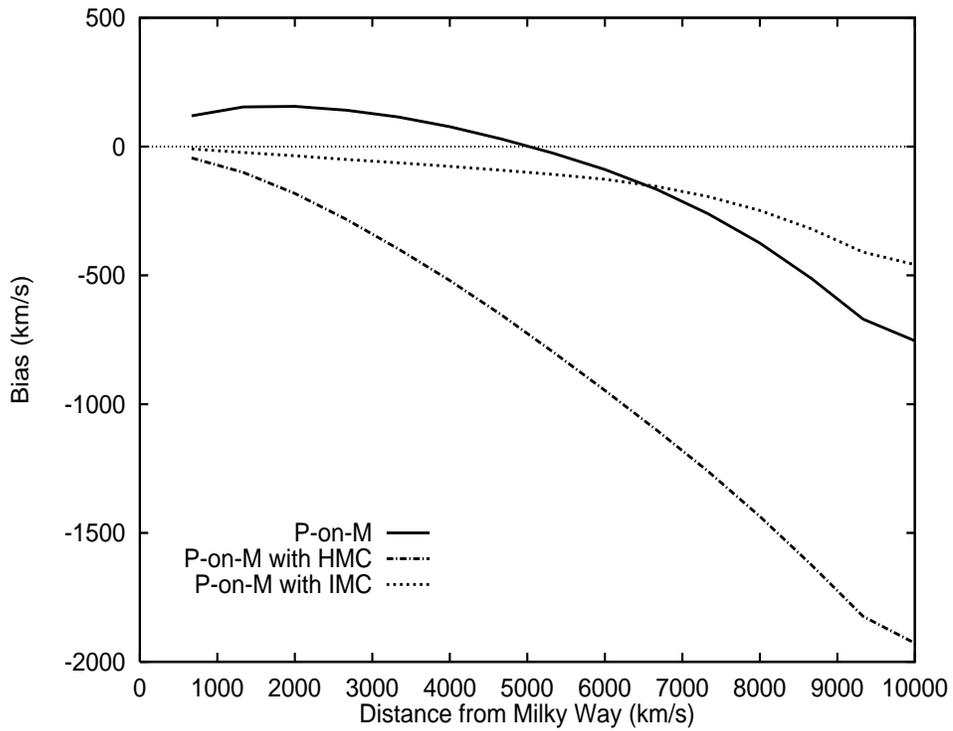

Figure 2: As in figure 1, but using the P-on-M estimator.

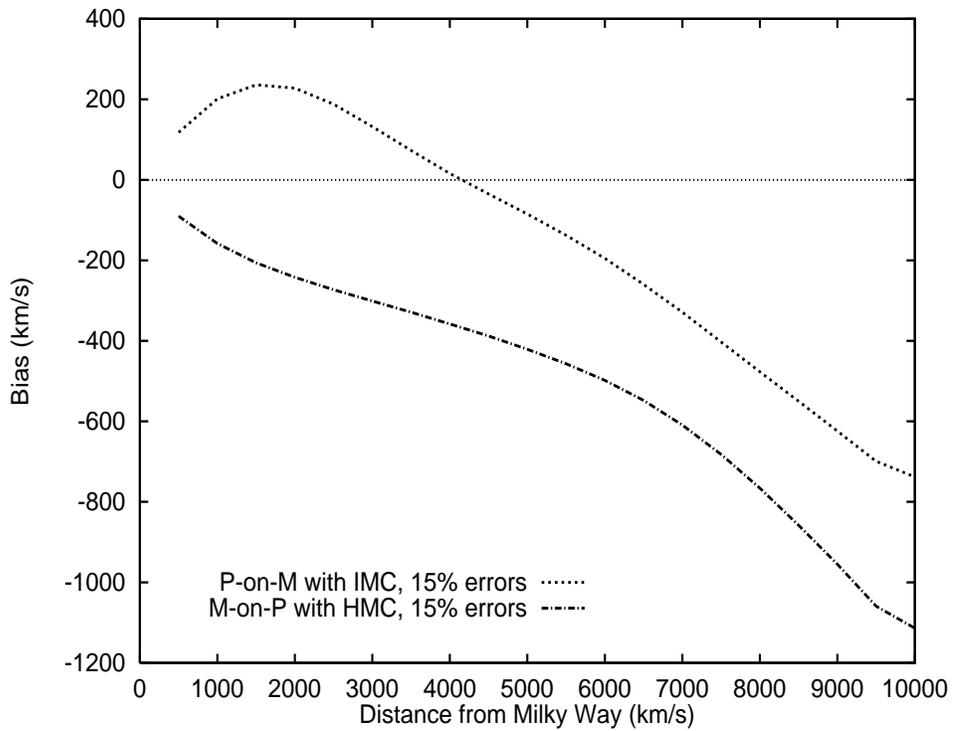

Figure 3: Bias as a function of distance for POTENT recoveries using distance estimators with log distance errors of about 15%.

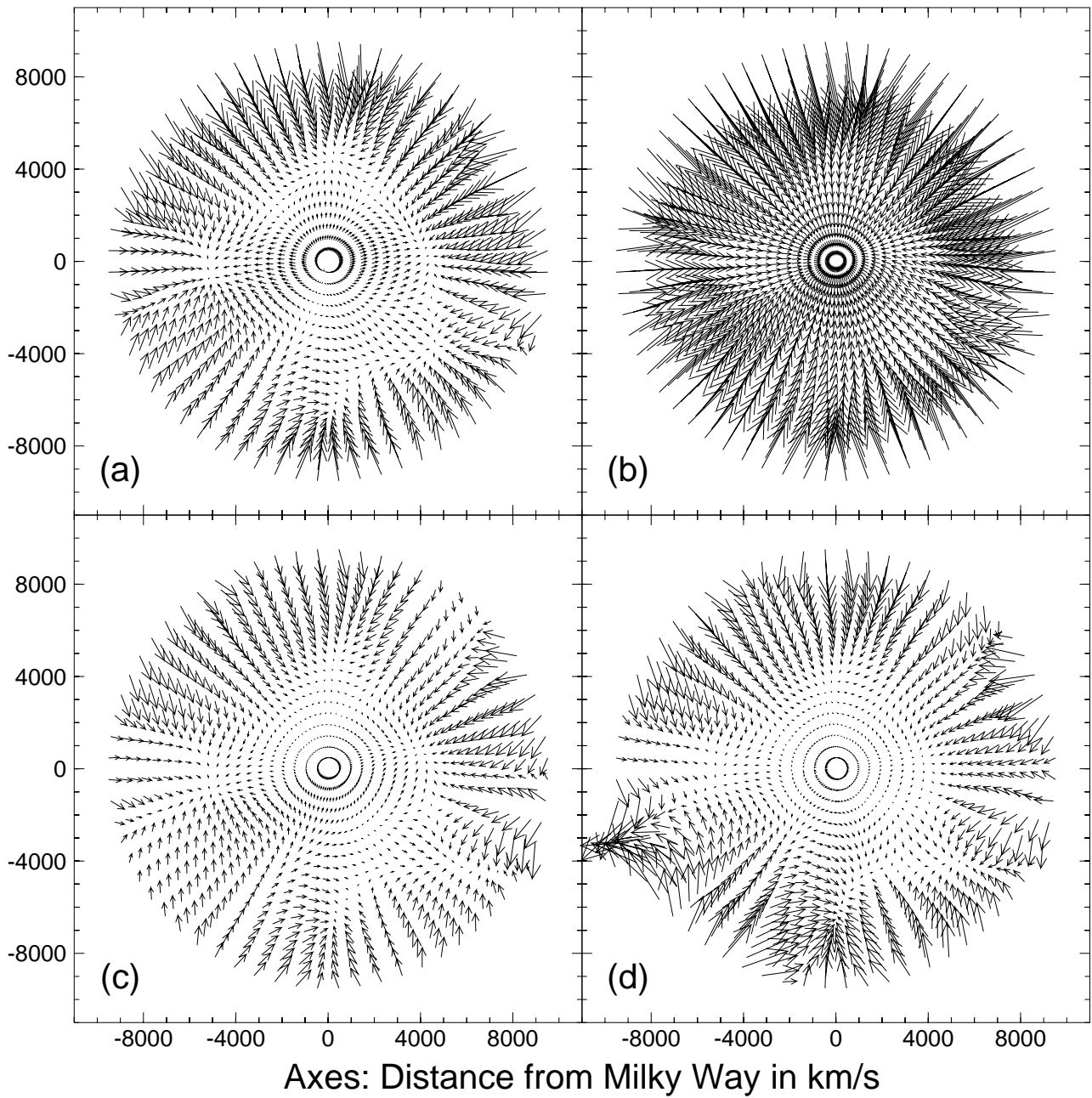

Figure 4: Four slices through Monte Carlo recoveries of quiet Hubble flow using POTENT on realisticly inhomogeneous data. Graph (a) show the recovery with 'raw' P-on-M estimates, (b) with the same estimates but an homogeneous Malmquist correction applied and (c) with an inhomogeneous correction. Graph (d) has an inhomogeneous correction calculated separately for each galaxy using a cone centered around it wide enough to contain 200 other galaxies.

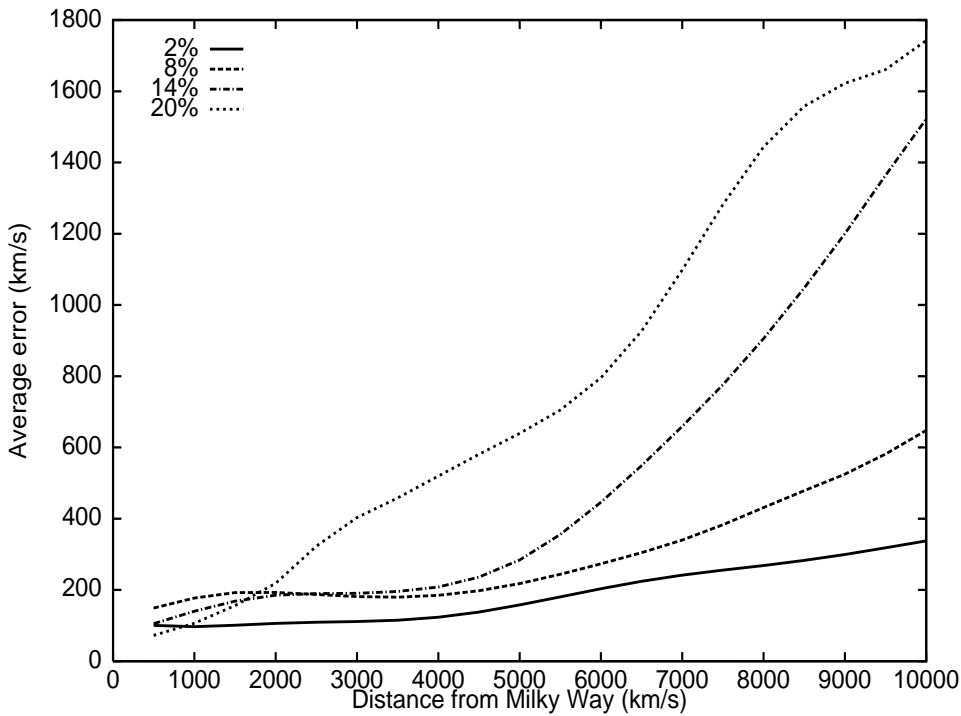

Figure 5: Errors in POTENT recoveries for various log distance errors. P-on-M estimator used with IMC applied.

Next, we need to consider a universe with inhomogeneities in the data from clusters, voids, obscuration and incomplete samples. To create such a sample, we use a combination of data from [10] and [3]. Figure 4 shows slices through four Monte Carlo recoveries of quiet Hubble flow. The first three (a, b and c) are the cases already considered for the P-on-M estimator and again show a clear advantage for the inhomogeneous Malmquist correction. However, as part of the formulation of the correction, it is necessary to average over the entire sky, giving the same correction for all parts of the sky. The complex residual biases seen in figure 4(c) are mainly due to this effect. Figure (d) is an attempt to improve on this by dividing the sky up into a number of cones, one centered on each galaxy and containing some fixed number of galaxies. The correction is then derived for each cone and applied to the one galaxy at the center. However, as can be seen from the figure, the results are not very promising. This is because in using only a sub-sample, noise in the histogram of distances used in the correction is increased and for cones narrow enough to make the method worthwhile, this noise is too great for reliable corrections. However, with a significant increase in sample size, this problem could clearly be overcome.

## 6.1 Varying the Log Distance Errors

Of course, minimising the bias is of little importance if the recovery is dominated by noise, so how good an estimator do we need to obtain suitably accurate peculiar velocity fields? Figure 5 shows how the error in the velocity field varies with both distance and error in log distance.

The errors are calculated by comparison to the biased velocity field and are, therefore, an accurate representation of the spread in the velocities. As before, the galaxies are drawn from an homogeneous universe and subjected to selection before having their distances estimated and corrected for velocity bias.

## 7 Conclusions

From these results we can see that the minimisation of bias from the recovered velocity field is of considerable importance, particularly as new surveys improve coverage to the extent that distant areas with correspondingly large distance errors start to produce seemingly useful results.

Given this, it is clear that the use of M-on-P estimators cannot be justified, with good results only obtained by a lucky coincidence of inappropriate corrections and log distance errors. However, P-on-M estimators fare little better if left to themselves. Correction of some sort is needed and the inhomogeneous Malmquist correction of Landy and Szalay is the best available as yet, despite its uncertain basis and its weakness towards the edge of the sample. However the fact that it is a correction only for radial inhomogeneities means that for realistic distributions of galaxies, its use must be carefully justified.

Overall, it seems that the velocity recoveries produced by POTENT are very sensitive to the introduction of biases. However, approximate corrections on the distance estimates used in the smoothing, such as IMC, are adequate if treated carefully. Also, gains in the accuracy of distance estimators hold the promise of considerable improvements in velocity field recoveries by a variety of methods.